# Deep Survival Analysis from Adult and Pediatric Electrocardiograms: A Multi-center Benchmark Study

Platon Lukyanenko,[1,2] Joshua Mayourian,[2,3] Mingxuan Liu[4], John K. Triedman,[2,3] Sunil J. Ghelani,[2,3] William G. La Cava[1,2,*]

[1]Computational Health Informatics Program, Boston Children's Hospital, Boston, MA, USA
[2]Department of Pediatrics, Harvard Medical School, Boston, MA, USA
[3]Department of Cardiology, Boston Children's Hospital, Boston, MA, USA
[4]National University of Singapore, Singapore
*Correspondence: william.lacava@childrens.harvard.edu

**ABSTRACT**

**Background**
Artificial intelligence applied to electrocardiography (AI-ECG) has recently shown potential for mortality prediction, but heterogeneous approaches and private datasets have limited generalizable insights into AI methodologies fit for this purpose. To address this, we systematically evaluated model design choices across three large medical center cohorts: Beth Isreal Deaconess (MIMIC-IV: n = 795,546 ECGs, United States), Telehealth Network of Minas Gerais (Code-15; n = 345,779, Brazil), and Boston Children's Hospital (BCH; n = 255,379, United States).

**Results**
We comprehensively evaluate models to predict all-cause mortality, comparing horizon-based classification and deep survival methods various neural architectures including convolutional neural networks and transformers. We also benchmarked against demographic-only and gradient boosting baselines. Top models yielded good performance (median concordance, Code-15: 0.83; MIMIC-IV: 0.78; BCH: 0.81). Incorporating age and sex improved performance across all datasets. Classifier-Cox models exhibited site-dependent sensitivity to horizon choice (median Pearson's R, Code-15: 0.35; MIMIC-IV: −0.71; BCH: 0.37). External validation reduced concordance, and in some cases, demographic-only models outperformed externally trained AI-ECG models on Code-15. However, models trained on multi-site data outperformed site-specific models by margins ranging from 5% to 22%.

**Conclusions**
These findings highlight several key factors for robust AI-ECG deployment. Deep survival methods consistently provided advantages over horizon-based classifiers, while inclusion of demographic covariates such as age and sex improved predictive performance across sites. The sensitivity of classifier-based models to horizon selection underscores the need for site-specific calibration. The multi-site experiment reveals that cross-cohort training, even between adult and pediatric cohorts, can substantially improve performance on those cohorts compared to cohort-specific training. Together, these results emphasize the importance of model type, demographic features, and training data diversity in developing AI-ECG models that can be reliably applied across populations.

**KEYWORDS**

Electrocardiography, AI-ECG, Convolutional Neural Networks, Survival Analysis

## INTRODUCTION

Electrocardiography (ECG) measures the electric activity of the heart. Abnormal 12-channel ECGs often indicate cardiovascular pathology and are thus a marker for disease and mortality. AI-ECG refers to the application of AI or machine learning to ECG interpretation. A common AI-ECG task is risk prediction, which can be framed as predicting event occurrence (e.g. mortality or passing a diagnostic threshold). Several studies have recently applied AI-ECG to predict patient outcomes including mortality risk [1–5], ventricular hypertrophy [6], and ventricular dysfunction in both adults [7] and children [8]. While these studies demonstrate AI-ECG's potential value, most use private data and only report a single, or a small set, of modeling approaches. Consequently, there is little consensus on best approaches to AI-ECG model development in general, including basic task formulation (e.g., classification or time-to-event prediction), the choice of model architecture, covariates, and other training procedures.

Most studies develop AI-ECG models on single, large, private ECG datasets (e.g., ECGs: 2.4M [9], 2.3M [4]; 1.2M [10]); the majority report results for a single AI-ECG model configuration (e.g. [1-2,9-10,15]), and still fewer make models publicly available (e.g., [1,15]). Another notable challenge in the mortality prediction setting is for site-specific models to generalize to new sites; Sau et al. [10] and Hughes et al. [1] report model performance drops (AUROC) of 6-20% in external testing.

Verifiably evaluating different modeling approaches and creating reproducible public ECG models requires large, public ECG datasets linked to high-quality patient-level data. Fortunately, such data has recently become more available with the public releases of the MIMIC-IV-ECG dataset (~800k ECGs) in 2023 [13-14] and the Code-15 dataset (~345k ECGs) in 2020 [15], vastly expanding on pre-2020 datasets [16].

Given the high interest in risk prediction from ECGs and the availability of these new datasets, the primary goal of this study is to benchmark state-of-the-art deep learning architectures and survival modeling approaches to inform AI-ECG best practices. To do so, we develop AI-ECG models of all-cause mortality using Code-15, MIMIC-IV-ECG, and a pediatric dataset from Boston Children's Hospital (BCH). We quantify model performance as a function of model architecture, survival analysis approach, and covariate inclusion choice, across datasets, classifier time horizons, and patient subgroups. We create robust baseline models to contextualize AI-ECG performance relative to traditional machine learning approaches and simple baselines. Finally, we evaluate the impact of cross-site training on model generalization and site-specific performance.

## BACKGROUND

The majority of AI applications to ECG signals have considered simpler prediction tasks than mortality; for example, many studies focus on arrhythmia detection, where classical AI methods leveraging feature extraction regularly report accuracies exceeding 99% [40].

Mortality prediction from ECGs is a more complex task, given the myriad pathologies that influence risk, yet has considerable utility from a clinical point of view, as it provides a general way for providers to risk-stratify patients and manage treatment. Recent studies have demonstrated 5-year mortality prediction from ECGs, beginning with Raghunath et al. in 2020 [4] and, more recently Hughes et al's work in 2023 [1]. Many studies utilize the time-series-adapted residual network that was proposed by Ribeiro et al in 2020 [4]. However, these works did not fully consider mortality as a time-to-event prediction (i.e., as a survival analysis problem), nor did they utilize state of the art approaches to deep learning for survival analysis [20]. Most recently, Sau et al [10] demonstrated survival analysis from ECGs using a deep survival approached called DeepHit [24]. However, the study does not provide code or sufficient methods details to support future research, which is one of our motivations here.

Our study covers these methods and benchmarks several additional approaches. For example, we benchmark a recent transformer-based method called ECGTransform [30], as well InceptionTime, which is considered a state-of-the-art time series classification architecture [26]. We benchmark three additional deep survival approaches. In contrast to prior studies, we evaluate the impact of cross-site training, benchmark on two large, publicly available databases, and make our code and model benchmarks publicly available to support downstream scientific research.

## METHODS

### *Patient Populations*

Code-15 [15] is an ECG dataset from the Telehealth Network of Minas Gerais, a Brazilian public agency providing telehealth services to Minas Gerais and Amazonian and Northeast states. Patient ECGs were recorded in primary care facilities by technicians and examined remotely by a cardiologist. The publicly available dataset includes 345,779 ECGs collected from April to September of 2018. Code15 ECGs are 7-10 second signals sampled at 400Hz, centered and padded with zeros to total a length of 4096. The Code-15 Dataset provides multiple ECGs per subject and indicates the patient's age and the follow-up time after the patient's final ECG. We only use the one entry per subject that provides a specific follow-up time.

MIMIC-IV-ECG [13], referred to as simply MIMIC-IV for brevity, includes 795,546 ECGs from 159,608 patients collected between 2008-2019 at the Beth Israel Deaconess Medical Center in Boston, Massachusetts. Patient ECGs were recorded in various settings, including emergency settings, hospitals, and outpatient care centers. Each ECGs is a 10 second signal sampled at 500Hz.
MIMIC-IV tracks date-of-death with state and hospital records, and censors deaths one year after a final recorded hospital visit.

BCH includes 225,379 ECGS from 79,568 patients collected 1990 to 2018 at Boston Children's Hospital in Boston, Massachusetts from emergency, hospital, and outpatient care settings[17]. In contrast to the adult cohorts, this dataset predominantly represents a pediatric congenital heart disease cohort. BCH ECGs are 8-second signals sampled at 250Hz. Death was tracked by an internal institutional database.

### *Dataset Preparation*

Dataset preparation is shown in **Figure 1**. Data is limited to [Patient ID, Time-To-Event, Event, ECG, Age, Sex, Machine Measures (MIMIC-IV only)], and entries with missing ECG samples or Time-To-Event are excluded. Time-To-Event is set to a minimum of 1 day. All ECGs are limited to 7 seconds, resampled to 400Hz, and symmetrically padded with zeros for an end shape of 12 x 4096. ECGs are split 64/16/20 into train/validation/test sets by patient ID. We do not limit the number of ECGs per patient. The random seed determines the Training/Validation split while the Test set remains fixed. Before model training or evaluation, all ECGs are z-score normalized per ECG channel based on the training set.

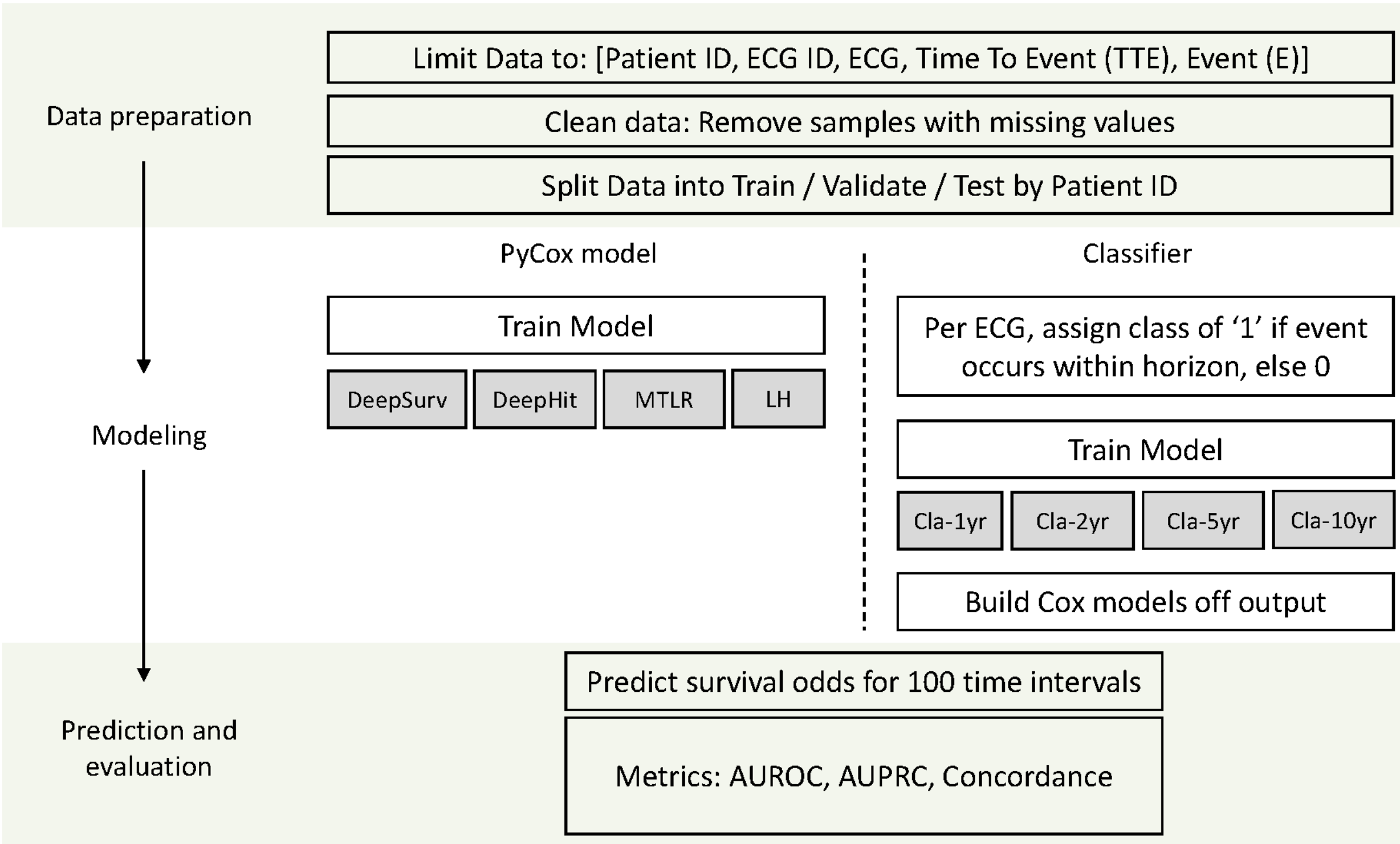

**Figure 1: Flow diagram describing the experimental design.**

***Survival Analysis***

Survival analysis builds survival functions, S(t), denoting the probability of not having experienced an event by a time t. Survival analysis uses data in the form [Time-To-Event, Event], which denotes event state at a follow-up time. This leverages information even when final event time is unknown: a device that works for two years before being lost still provides two years of evidence of non-breakage. 'Censoring' refers to not knowing event outcomes for some subjects.

The most common survival functions are Kaplan-Meier curves which display survival over time from a tracked population. If the population is clustered into groups, an individual's trajectory can be estimated from their cluster's survival function. Otherwise, survival functions usually fit an exponential decay or Weibull function to regressors (demographics, measurements, etc.).

The Cox proportional hazards model [18] models hazard, $h(t) = d/dt\ (1 – S(t))$, as an unknown base function scaled by exponential decays: $h(t) = h_0(t)\exp(\beta_1x_1+\beta_2x_2+...+\beta_mx_m)$ where $\{x_1, ..., x_m\}$ are regressors and $\{\beta_1, ..., \beta_m\}$ are learned weights. The Breslow estimator is often used to fit $h_0(t)$ [19].

The Cox regression assumes that *log-risk* $(\beta_1x_1+\beta_2x_2+...+\beta_mx_m)$ is *proportional* (e.g. smoker risk / non-smoker risk = constant) and a *linear* function of regressors that is *time-constant* (there are no time-regressor interactions).

Broadly, there are two survival modeling approaches with neural networks:

***Classifier-Cox (CC)***: The first approach converts inputs into a set of values through a neural network. These values are then treated as generic markers to fit a Cox regression [18]. The neural networks are typically trained to return a single value classifying inputs by whether an event occurs within a time horizon or not. This approach is easily implemented and can account for new covariates in the Cox regression stage but must handle data censoring when assigning classifier labels.

***Deep Survival (DS)***: The second approach trains neural networks with survival-specific loss functions to directly generate survival curves. We evaluate four different Deep Survival approaches from the PyCox[20] package. DeepSurv [21] models *log-risk* as a *time-constant, proportional, non-linear* function. LogisticHazard (LH) [22] models *log-risk* as a *time-varying, non-proportional, non-linear* function and attempts to eliminate batch size effects. MTLR [23] models *risk* with a *time-varying, non-proportional, logistic regression* on a *non-linear* function. DeepHit [24] models *risk* as a *time-varying, non-proportional, non-linear* function and allows for several competing risks – i.e., multiple possible outcome events – which we do not exploit in this study.

### *AI-ECG survival models*

Several recent studies predict mortality from ECGs. These are either built completely on large, private, ECG banks [3,4,5,10,16,25] or are fine-tuned in the context of a particular setting or population[2,11,12]. Some studies use Deep Survival[3,10,11], and some use Classifier-Cox approaches [4,5,25]. Most studies use a convolutional network to interpret ECG, with a slightly-more-popular choice being the ResNet architecture from Ribeiro et al.[5,10,17,25].

### *Benchmark Procedure*

Benchmark parameters are in **Table 2**. In total, more than 3,000 AI-ECG models are analyzed. We use two convolutional neural network (CNN) architectures, InceptionTime [26] and a modified ResNet architecture [15], and one transformer architecture, ECGTransform [30]. For each architecture, we train four deep survival models[21] and four classifier-Cox models where neural net classifiers predict mortality up to one, two, five, and ten-year horizons. Models are trained with and without age and sex covariates. Additional modeling details are in the Supplement. We make this resource publicly available (see Data and Resource Availability Section).

**Table 1: Settings for the mortality prediction benchmark experiment.**

| Setting | Values |
|---|---|
| **Survival Analysis** | |
| **Deep-Survival** | DeepHit, DeepSurv, LogisticHazard, Multi-Task Logistic Regression |
| **Classifier-Cox** | Classification mortality time horizon (years): 1,2,5,10 |
| **AI-ECG Architecture** | InceptionTime (CNN), Resnet (CNN), ECGTransformer |
| **Covariate options** | None, Age+Sex, Age+Sex+ECG Machine Measure (MIMIC-IV only) |
| **Demographic-only models** | XGBoost, Feedforward Network |
| **Normalization** | z-score per channel, based on model's training data |

### *Neural Network Architectures*

Our neural networks had three modules: an ECG-processing module, a covariate-processing module (three feedforward-ReLU layers with output dimension 32), and a fusion module connecting the other two pieces (three feedforward-ReLU layers with output dimension 128) that heads into a final linear layer (dimension 1 for Classifier-Cox models and DeepSurv, dimension 100 for LH, MTLR, and DeepHit – one per time bin). When covariates are not included, the covariate-processing module is skipped. When ECG is excluded for covariate-only baselines, the ECG module returns a one-dimensional value of ‘0’, resulting in a feed-forward network.

Along with the survival approaches, we benchmark three neural network architectures. “ResNet” is a multi-channel time series [15,25] adaptation of the original ResNet [27]. Architecture parameters were kept at defaults tuned to the full Code dataset. This model has 6.9-7.5M parameters. “InceptionTime” [26] adapts AlexNet [28] by widening convolutional kernel widths and including channel-wise bottlenecks to control model complexity. This architecture performs well on many small time-series classification benchmarks and has recently been used in fetal heart rate monitoring [29]. Architecture parameters were kept at publication defaults (kernel widths 11, 21, 41). This model has 510-530k parameters. “ECGTransForm” is a recent transformer architecture combining multi-scale convolutions with a bidirectional transformer. It reports state-of-the-art performance in arrhythmia classification and has 2.6M parameters [30].

### *Baselines*

We generate baseline models by training XGBoost and feedforward networks (‘FF’) on age and sex covariates. In addition, we leverage the automated machine measures available in MIMIC-IV (Axes – P, QRS, T; Durations – P, PQ, QRS, QT, RR.). We generate a baseline model using these measures, along with age and sex, as a baseline feature extraction modeling approach. We additionally test providing measures to AI-ECG models.

### *Outcome Measures and Statistics*

Our primary outcome measure is the Concordance Index (concordance). Like area under the receiver operating characteristic curve (AUROC), concordance evaluates the ability of a model to rank patients correctly by risk, but also takes into account the timing of the event. At the time of an event, a subject should be at a higher risk than any other subject still under observation; concordance is the fraction of correctly ranked subject-subject comparisons. Additional measures (AUROC, AUPRC, time-censored concordance) are available in the Supplemental Material.

Unless otherwise mentioned, paired comparisons use the Wilcoxon test and unpaired comparisons use the Mann-Whitney test. Multiple comparisons are adjusted for with Benjamini-Hochberg corrections at a 5% false discovery rate.

## RESULTS

### *Patient Populations*

The patient cohorts resulting from our data preparation and inclusion criteriea are summarized in **Table 1**. Overall, our experiment is conducted over 1,200,658 ECGs from 443,216 patients. For Code-15, we analyze 233,647 ECGs; 1.2%, 2.1%, 3.4%, and 3.6% of ECGs link to a mortality by year 1, 2, 5, and 7.7 (the max follow-up time), respectively. For MIMIC-IV, we kept 785,035 ECGs; 14.8%, 18.4%, 24.5%, 27.6%, and 27.8% of ECGs link to a mortality by year 1, 2, 5, 10, and 13, respectively. For BCH, we kept 181,976 ECGs; 0.9%, 1.5%, 3.1%, 4.8%, and 6.9% are associated with a mortality by year 1, 2, 5, 10, and 33, respectively.

**Table 2*: Population characteristics for the three sites used for benchmarking.*** Note that age and event rates differ between sites. Categorical p-value comparisons use the Chi-square test; numerical comparisons use Student's t-test.

| Site | Measure | | All | Event (Mortality) | Non-Event | p value |
|---|---|---|---|---|---|---|
| **Code-15 (2018)** | ECGs, N (%) | | 233,647 | 8,341 (3.6) | 225,306 (96.4) | |
| | Patients, N (%) | | 233,647 | 8,341 (3.6) | 225,306 (96.4) | |
| | Sex, N (%) | Female | 138,911 (59.4) | 3,853 (2.8) | 135,058 (97.2) | << 1E-5 |
| | | Male | 947,36 (40.6%) | 4,488 (4.7) | 90,248 (95.3) | |
| | Mean Age, Yr (Std) | | 50.7 (19.8) | 70.4 (14.0) | 50.0 (19.6) | << 1E-5 |
| | Mean Follow-up Time, Yrs (Std) | | 3.7 (1.9) | 2.0 (1.6) | 3.7 (1.8) | << 1E-5 |
| **MIMIC-IV (2008 - 2019)** | ECGs, N (%) | | 785,035 | 215,039 (27.4) | 569,996 (72.6) | |
| | Patients, N (%) | | 159,122 | 25,107 (15.8) | 134,015 (84.2) | |
| | Sex, N (%) | Female | 384,923 (49.0) | 99,170 (25.8) | 285,753 (74.2) | << 1E-5 |
| | | Male | 400,112 (51.0) | 115,869 (29.0) | 284,243 (71.0) | |
| | Mean Age, Yrs (Std) | | 62.3 (17.1) | 70.4 (14.2) | 59.2 (17.1) | << 1E-5 |
| | Mean Follow-up Time, Yrs (Std) | | 2.6 (2.6) | 2.0 (2.4) | 2.9 (2.6) | << 1E-5 |
| **BCH (1990 – 2018)** | ECGs, N (%) | | 181,976 | 12,638 (6.9) | 169,338 (93.1) | |
| | Patients, N (%) | | 50,447 | 16,71 (3.3) | 48,776 (96.7) | |
| | Sex, N (%) | Female | 86,141 (47.3) | 6,048 (7.0) | 80,093 (93.0) | 0.23 |
| | | Male | 95,835 (52.7) | 6,590 (6.9) | 89,245 (93.1) | |
| | Mean Age, Yrs (Std) | | 12.8 (11.9) | 21.5 (16.7) | 12.2 (11.2) | << 1E-5 |
| | Mean Follow-up Time, Yrs (Std) | | 8.9 (6.7) | 7.2 (6.1) | 9.0 (6.7) | << 1E-5 |

### *Local Model Concordance*

The concordance of site-specific models, across model configurations, input and datasets, is summarized in **Figure 2**. Because transformers showed markedly worse performance, their results, along with additional baselines, are detailed in **Table S1**. The following analysis, statistics, and discussion focus on the CNN and competitive baseline models.

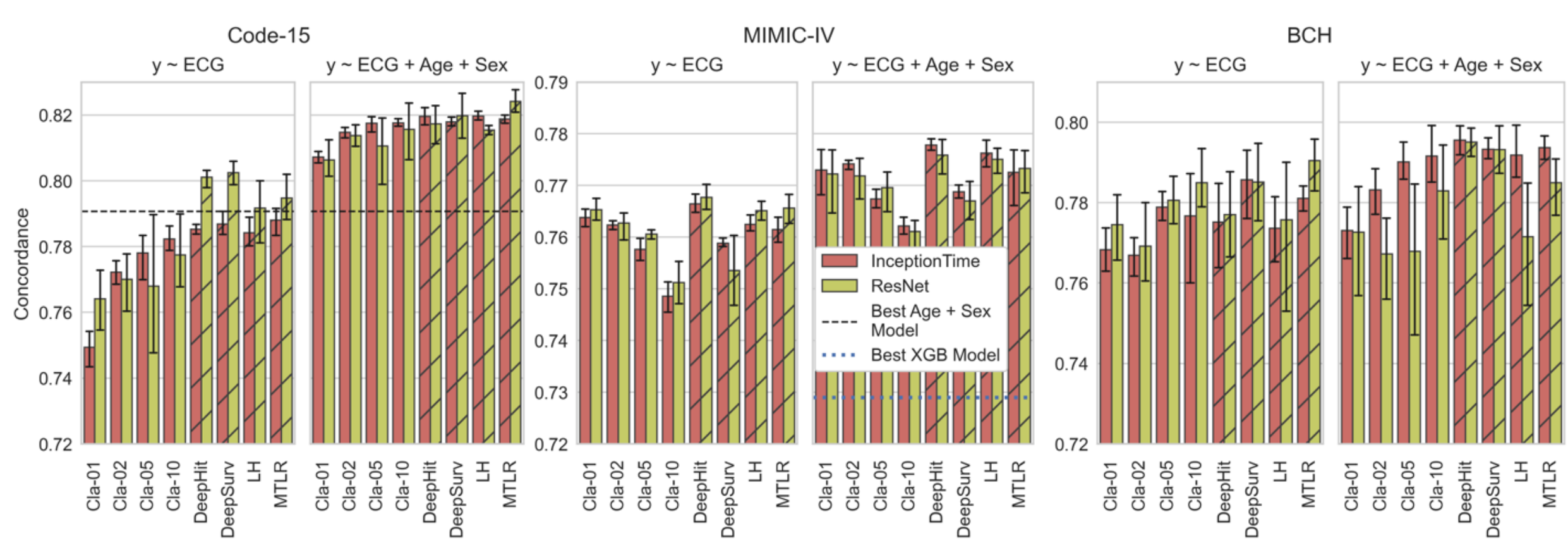

**Figure 2. Model concordances across datasets (subplots), covariates (columns), survival modeling approaches (x-axis), and architectures (color).** Hatched bars indicate deep survival approaches. "Cla-X" marks X-

year horizon Classifier-Cox models. “Best XGB Model” is a Cla-2 model trained on demographics and automatic ECG machine measures. Error bars denote 95% confidence intervals of mean concordance over 5 random seeds. Error bars denote 95% confidence interval.

The model configurations with best test set performance on each dataset are shown in **Table 3,** including performance at the 1 and 5 year marks. For cross-reference, chance mortality rates per year are in **Table S2.** For MIMIC-IV-ECG, appending automatic machine measurements as covariates did not noticeably improve model performance.

**Table 3: Concordance and AUPRC of top-performing, site-specific models, per dataset and input configuration**. For cross-reference, mortality rates per year are in Table S2.

| Dataset | Model Inputs | Approach | Architecture | Concordance | 1 Year AUPRC | 5 Year AUPRC |
|---|---|---|---|---|---|---|
| Code15 | y ~ ECG | DeepSurv | ResNet | 0.80 (0.80 - 0.81) | 0.07 (0.07 - 0.07) | 0.14 (0.14 - 0.14) |
| | y ~ ECG + Age + Sex | MTLR | ResNet | 0.83 (0.82 - 0.83) | 0.07 (0.07 - 0.07) | 0.14 (0.14 - 0.15) |
| MIMIC-IV | y ~ ECG | DeepHit | ResNet | 0.77 (0.77 - 0.77) | 0.42 (0.42 - 0.42) | 0.49 (0.49 - 0.49) |
| | y ~ ECG + Age + Sex | DeepHit | InceptionTime | 0.78 (0.78 - 0.78) | 0.45 (0.44 - 0.45) | 0.52 (0.51 - 0.52) |
| | y ~ ECG + Age + Sex + Machine Measures | DeepHit | ResNet | 0.77 (0.77 - 0.78) | 0.43 (0.43 - 0.44) | 0.46 (0.45 - 0.48) |
| BCH | y ~ ECG | MTLR | ResNet | 0.79 (0.79 - 0.80) | 0.05 (0.04 - 0.05) | 0.13 (0.13 - 0.14) |
| | y ~ ECG + Age + Sex | DeepHit | ResNet | 0.79 (0.79 - 0.80) | 0.05 (0.05 - 0.06) | 0.14 (0.13 - 0.14) |

### *Classifier-Cox performance is sensitive to time horizon*

We find that classifier-Cox horizon choice (the time at which binary mortality is determined) correlates with median model concordance in all cases (**Table 4**). This correlation can be positive (Code-15, BCH) or negative (MIMIC-IV), hindering the selection of a generally appropriate value.

**Table 4: Classifier-Cox models are sensitive to the choice of time horizon.** Pearson’s R correlation between classification model concordances and the choice of time horizon used in training. The direction of the effect is positive in BCH and Code-15 and negative in MIMIC-IV.

| | ECG + Demographic | | ECG Only | |
|---|---|---|---|---|
| Dataset | Pearson R | p value | Pearson R | p value |
| Code-15 | 0.35 | 2.66E-02 | 0.41 | 8.89E-03 |
| MIMIC-IV | -0.71 | 2.66E-07 | -0.88 | 1.08E-13 |
| BCH | 0.37 | 1.95E-02 | 0.47 | 2.35E-03 |

### *Deep Survival approaches outperform Classifier-Cox approaches*

Among local models trained with ECG and covariates, Deep Survival (DS) models, overall, show statistically higher concordance than Classifier-Cox (CC) models across sites ($p < 1.98E-03$, **Table S3**), although the performances differences are seldom large. On cross-evaluation, DS outperforms CC in 4/6 cases and CC outperforms DS in 1/6.

### *CNN model architectures perform similarly, and outperform Transformer architectures*

For models trained with ECG and covariates, InceptionTime (IT) outperforms ResNet (RN) on the BCH dataset; on cross-evaluation, RN outperforms IT in 3/6 cases and IT outperforms RN in 1/6 cases (**Table**

**S4**). Transformers, even on top individual runs (concordance, Code-15: 0.793, MIMIC-IV: 0.762, BCH: 0.774) do not exceed median Deep Survival CNN model concordance (**Table S1**).

***Including demographic covariates improves model performance***

In paired evaluations, AI-ECG models that included age and sex covariates always outperform ECG-only models ($p < 2.05E-4$, **Table S5**; **Figure 2**), improving concordance an average of 5.2%, and as much as 11.7%.

***Multi-site model training improves both global and local model performance***

**Figure 3** shows the performances of ResNet AI-ECG models in local validation and external validation across sites. As expected, models perform best when trained, at least partially, on data generated from the same population as their respective test sets. Covariate-only models (age and sex models) provide surprisingly strong baselines and generalize well between sites. This is especially noteworthy for Code-15: covariate-only models generalize to Code-15 cohort better than CNNs trained on other single site data.

The rightmost column of **Figure 3** also shows the performance of ResNet models trained on data from all sites. These provide maximum, or near-maximum, performance on all datasets, across modeling approaches. In **Figure 4**, we more directly compare the performance of locally and globally trained ResNet models on each site, which demonstrates that globally trained models outperform site-specific models, even on test data from the sites on which the local models are trained.

We investigate site-specific model performances across test cohorts on age and sex subgroups in **Figure S4.** Although cohorts differ by age and sex which may make generalization challenging, we do not find clear evidence that performance differences are explained by these age or sex differences.

***Survival models capture expected survival rates***

We compare Kaplan-Meier curves for Code-15 and MIMIC-IV to mean predicted population survival from ECG-only models in **Figure S1,** confirming that model estimates closely match expected survival rates throughout the study period.

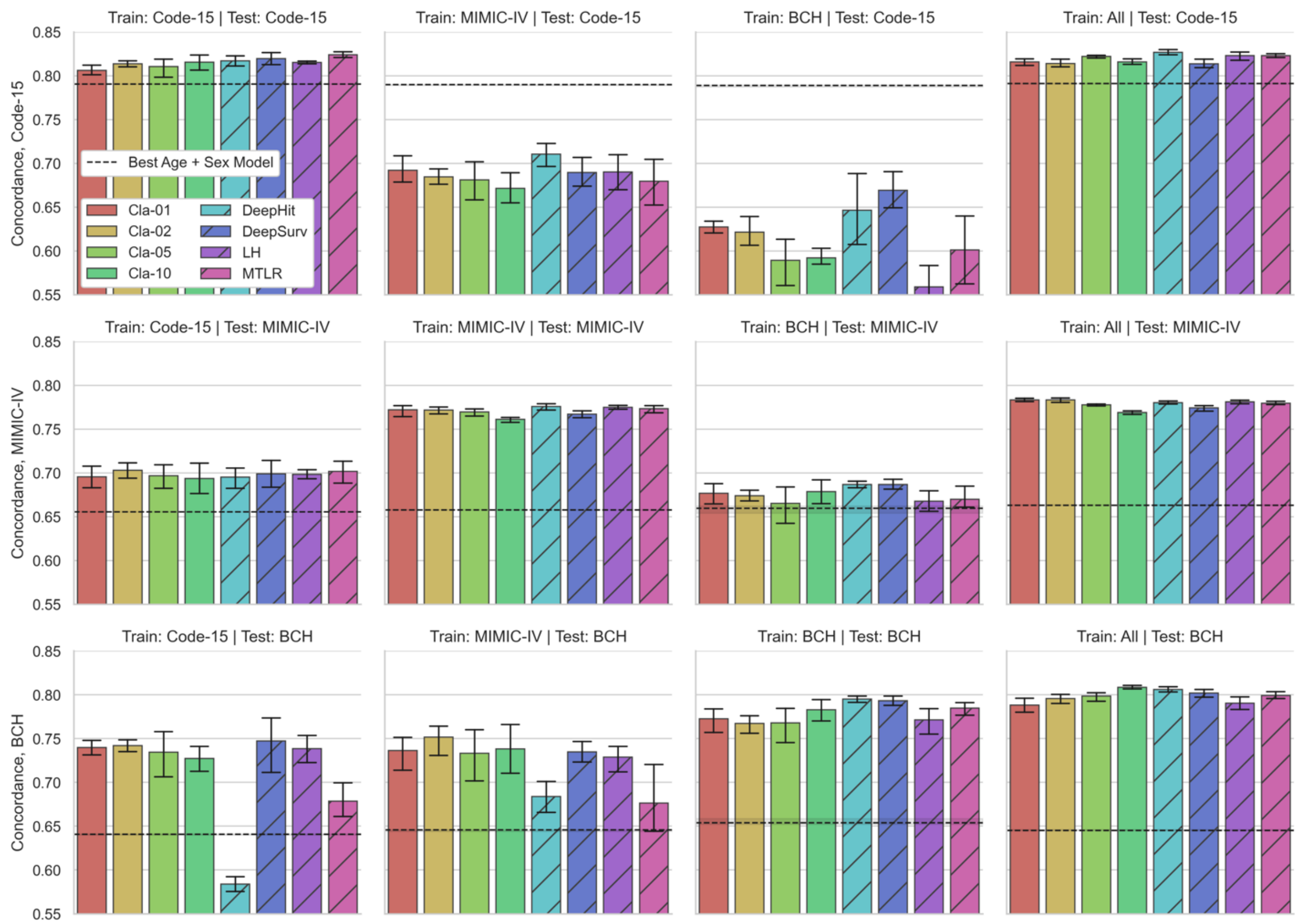

**Figure 3: Model concordance in cross-evaluation.** Rows denote test performance on Code-15, MIMIC-IV, and BCH, respectively. Columns denote which site the models were trained on. The rightmost column shows concordance for models trained across the training sets of all three sites. Bar color denotes the survival analysis approach; hatches denote deep survival approaches. The performance of the best demographic only baseline model (trained on age + sex) is shown by a dotted line. Results here are for the ResNet model architectures trained with ECGs and demographics. Error bars denote 95% confidence interval.

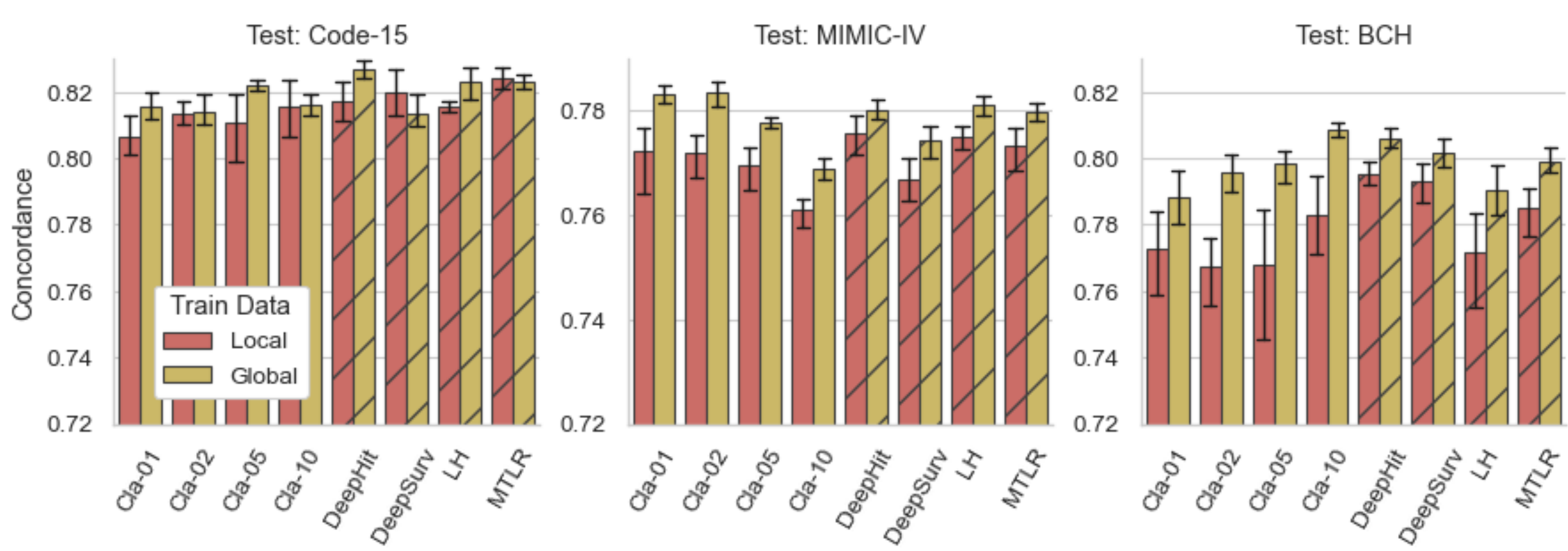

**Figure 4: Multi-site training improves local performance.** Median concordance of mortality models trained on site-specific data (light brown) versus training data from each site (red). Subplots denote the site-specific performance of these models, from left to right, on data from Code-15, MIMIC-IV-ECG, and BCH, respectively. X-axis denotes the modeling approach; all approaches use the ResNet architecture and include demographic covariates. Error bars denote 95% confidence interval.

### *Model explanations exhibit salient risk factors for mortality*

We investigate model explainability with a median-waveform analysis [8] in **Figure 5**. The median high-risk waveforms show lower amplitude and QRS complexes, as well as flattened/inverted lateral precordial T waves indicating several pathological findings (delayed myocardial activation, possibly myocardial strain). The QRS complex is most salient, suggesting focus mostly on myocardial activation (heartbeat dynamics). High salience of the V2 lead indicates anteroseptal activation and aligns with analyses of left ventricular (LV) dysfunction.Model explainability across various heart lesions for a ResNet Classifier-Cox approach is also explored in prior work [17].

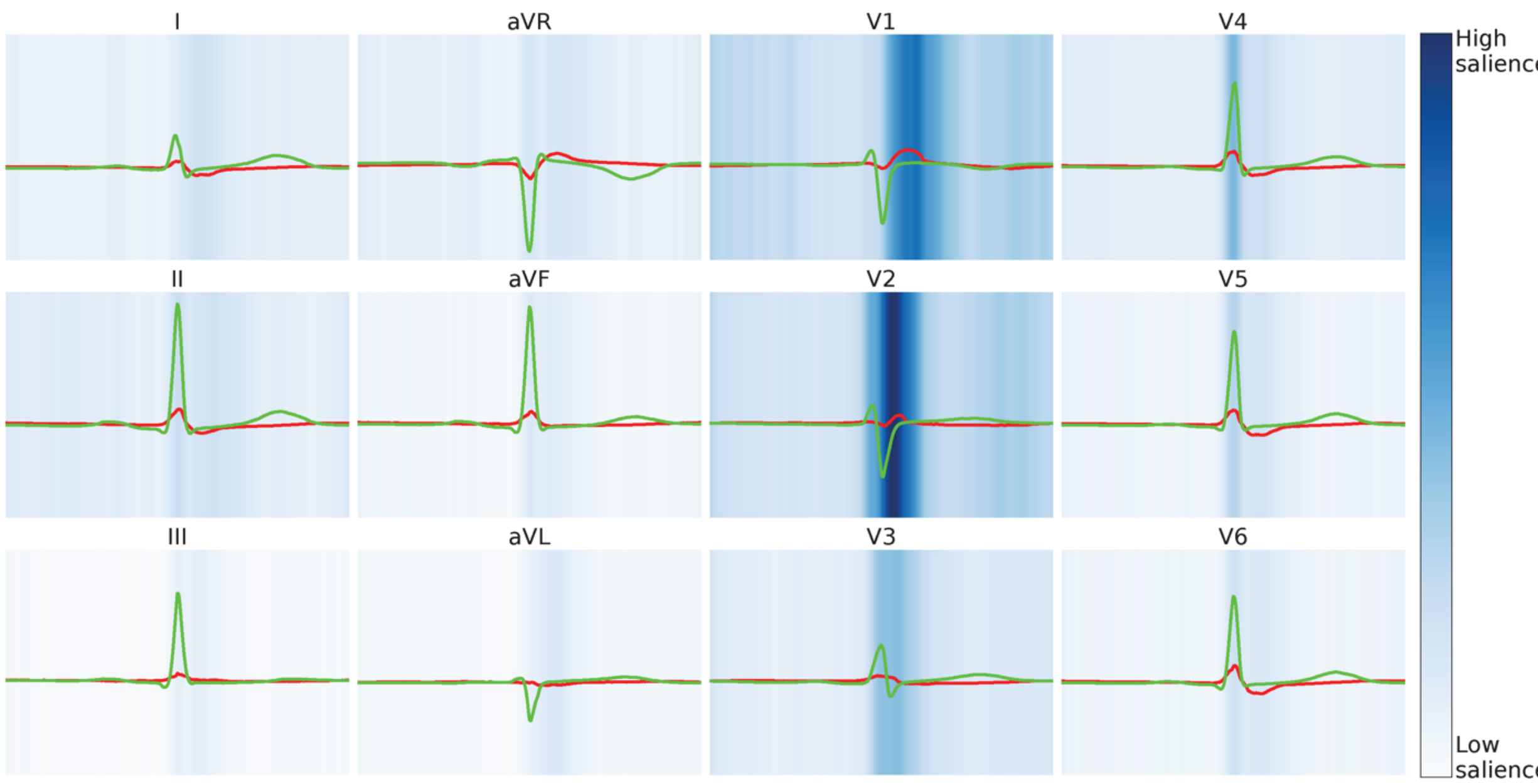

**Figure 5: Median heartbeats across 12-lead ECG and SHAP analysis for a MIMIC-IV DeepHit ResNet**. Green: median heartbeat for 100 ECGs with lowest-risk-estimate (mean prediction probability: 0%). Red: median heartbeat for 100 ECGs with highest-risk-estimate (mean prediction probability: 69%). Blue shading: SHAP salience. Model explanations highlight dampened QRS activity, indicative of LV dysfunction, in high-risk patients. For details of median waveform analysis, see [8].

## DISCUSSION

### *Contribution*

This work is the first to benchmark AI_ECG survival modeling approaches in all-cause mortality prediction on three broad datasets: Code-15, MIMIC-IV, and BCH. We provide verifiable results and baselines that inform ECG-processing and survival-analysis approaches for future AI-ECG modeling. We also release our code, which can be adapted to evaluate new modeling approaches or used to train AI-ECG survival-analysis models. This can also be applied beyond mortality modeling to model event occurrence (e.g. first diagnostic threshold crossing) from data in the [time-to-event, event] format. Our work may support future studies that develop, implement, and test AI-ECG models that prospectively inform patient care and case management.

### *Deep Survival approaches outperform Classifier-Cox approaches*

In this study we find Deep Survival approaches work better than Classifier-Cox approaches in two respects. First, the median Deep Survival models significantly increase model concordance over Classifier-Cox models. Second, deep survival approaches do not require a time horizon to be specified, and classifier-cox models exhibit performance sensitivity to this horizon choice. Our results indicate that horizon choice is both dataset-specific, with opposing trends for Code-15 and MIMIC-IV, and can substantially impact performance with the wrong choice. For example, short-horizon data includes few positive-event samples but assumptions about censored patients have little effect. Meanwhile, long-horizon data has more positive-event samples but the assumptions about censored patients have additional time to compound. Furthermore, when training classifiers it is necessary to combine (Time-to-Event, Event) data into a single variable to train a classifier (we discuss our approach in the Supplement**).** This choice can have substantial effects on what data models use (e.g. MIMIC-IV data is censored to 1 year unless linked to follow-up measures or a mortality). Deep Survival modeling does not require such a choice.

### *Comparisons to past work and model generalizability*

Our results confirm the performance of ECG mortality risk prediction models from past studies at the one-year mark (0.81-0.85 concordance) [4,5] and five-year mark (concordance: 0.78-0.83) [1,5]. Additionally, our ResNet DeepHit, ECG-only MIMIC-IV results (concordance: 0.77) concur with a recent study that trained a ResNet DeepHit model, dubbed “AIRE” [10], on 1.2M, 10-second ECGs from Beth-Israel Deaconess Medical Center in Boston, MA (concordance: 0.775). The results stand to reason as we expect this dataset to largely overlap with MIMIC-IV.

Prior work developing AI-ECG mortality prediction models reports performance deterioration of different degrees across sites, from 0-6% AUROC differences (SEER, [1]) to 1-18% concordance drops (AIRE, [10]). Follow-up evaluations of SEER indicate larger performance deterioration on the UK Biobank cohort (UKB AUROC 0.57 [10], vs 0.83 in Stanford cohort [1]). Our results confirm that AI-ECG models trained on site-specific data struggle to generalize to new sites, across AI-ECG models, and further suggest that these differences are not fully explained by demographic differences between cohorts (**Figure S2**) nor mortality rate differences (**Table S2**). However, by training on multiple sites, our study demonstrates that performance improves across cohorts and, importantly, outperforms site specific models.

### *The importance of baseline models and demographic covariates*

Past work rarely compares AI-ECG to simpler models, such as demographic-only models, as we do here, yet AI-ECG’s value depends on its favorable performance to less complex baselines. Our results have important implications for studies reporting generalization performance on Code-15 [15], where demographic-only baseline models (median concordance: 0.79) outperform the AI-ECG models tested here, as well as those in prior work (e.g. AIRE concordance: 0.76 [10]).

We conjecture that AI-ECG models may have more value in acute care settings (i.e., MIMIC-IV, BCH) than in outpatient settings (Code-15) due to the clinical context. In general, our results strongly support the include of age and sex covariates in all AI-ECG mortality prediction models.

### *Limitations*

While we conducted broad sweeps of model configurations in this study, they are not exhaustive. We cannot rule out that other model architectures, or fine-grained tuning of the architectures we benchmark, may improve performance on specific datasets. Furthermore, our results are specific to mortality risk prediction from AI-ECGs, which is one of several prediction targets of interest. Future work could expand the clinical interpretation of model behavior to understand specific pathologies leading to mortality, which we do not inspect deeply here. As a retrospective benchmark, this study may not fully capture deployment considerations; the ethical deployment of survival models in a clinical setting requires that model performance be evaluated in those contexts, with appropriate oversight and a focus on how model use impacts patient care in specific clinical workflows.

### *Conclusions*

Based on this study, we recommend the following for AI-ECG model development: 1) including demographic covariates; 2) using deep survival, rather than classification, approaches; 3) the 1-D ResNet architecture; 4) supplementing local model training with data from additional sites. These model choices generally provide better performance, avoid classifier difficulties related to horizon choice and data censoring, and rely on a well-vetted architecture. Taken together, these design choices can be responsible for most of the improvements AI-ECG offers over simpler baseline models, which we measure to be 5-25% median concordance improvement, depending on cohort. We hope this study and its recommendations provide a valuable resource for future ECG and ECG-mortality studies.

## RESOURCE AVAILABILITY

### *Lead contact*

Requests for further information and resources should be directed to and will be fulfilled by the lead contact, William G. La Cava (william.lacava@childrens.harvard.edu).

### *Materials availability*

No materials were generated in the study.

### *Data and code availability*

#### Data

- This paper analyzes existing, publicly available data, accessible at:
  - MIMIC-IV-ECG: https://doi.org/10.13026/4nqg-sb35 under the Open Data Commons Open Database License v1.0
  - Code-15: https://doi.org/10.5281/zenodo.4916206 Creative Com-
  - mons license CC-BY 4.0
- The BCH data reported in this study cannot be deposited in a public repository because of Institutional Review Board policies to protect pediatric patient privacy. To request access, contact the lead contact above.

#### Code

- All original code has been deposited at https://github.com/cavalab/ecg-survival-benchmark and is publicly available under the Gnu Public License v3.
- The top deep survival models for MIMIC-IV and Code-15 cohorts are available on Zenodo at DOI 10.5281/zenodo.16877773.

#### Additional information

Any additional information required to reanalyze the data reported in this paper is available from the lead contact upon request.

## ACKNOWLEDGMENTS

We would like to thank Lorenzo Peracchio for help with data validation. P.L., and W.G.L. were partially supported by the Kostin Innovation Fund at Boston Children's Hospital. W.G.L. was partially supported by National Institutes of Health grant R01LM014300.

## AUTHOR CONTRIBUTIONS

Conceptualization, P.L., J.M., W.G.L.; methodology, P.L., W.G.L.; investigation, P.L., W.G.L.; writing—original draft, P.L., W.G.L.; writing—review & editing, J.M., S.G., J.K.T., W.G.L.; funding acquisition, J.K.T.; resources, W.G.L..; supervision, J.M.,J.K.T., S.G., W.G.L..

## DECLARATION OF INTERESTS

None declared

## DECLARATION OF GENERATIVE AI AND AI-ASSISTED TECHNOLOGIES

AI and AI-assisted technologies were not used in preparing this manuscript.

## SUPPLEMENTAL INFORMATION

**Document S1. Supplemental methods, Figures S1–S2, and Tables S1–S6**

# SUPPLEMENTAL INFORMATION

## ADDITIONAL MODELING DETAILS

For full training details and code, refer to the Github repository: https://github.com/cavalab/ecg-survival-benchmark .

### *Model Training*

Models train for up to 200 epochs, stopping early if their performance on the validation set does not improve for 20 consecutive epochs. As in past work[31], we use the AdamW optimizer, start with a 1e-3 learning rate, and scale the learning rate by 0.1x if validation performance does not improve for 10 consecutive epochs to a minimum of 1e-8[32]. We trained models with five random, where the seed also determines the training/validation data split. Models were trained on a cluster that unpredictably provided a A40, A100, L40 or Nvidia Quadro GPU. Data was loaded to RAM, requesting 100GB for Code-15, 150GB for BCH, and 300GB for MIMIC-IV. Training was done on batches of 512 shuffled data samples with 256 loaded onto a GPU concurrently. Per configuration, we evaluate the model with minimum validation loss.

Additionally, DeepSurv requires GPU batches to contain positive events for loss calculations; for these models, GPU-batches that contained no positive-event samples had their last sample replaced with a positive-event sample chosen randomly from the training set.

Code-15 models averaged 4.5 hours to train and achieved maximum validation performance by 30 epochs. MIMIC-IV models took 7.2 hours and achieved maximum validation performance by 12 epochs. Training times depend on GPU, with faster GPUs taking 600s/epoch and slower ones up to 2200s/epoch. GPU batch sizes were based on slower GPUs.

When Test and Train dataset lead orders did not match (Code15 differs from MIMICIV), the Test dataset was reordered to match the Train dataset's format.

### *Survival Modeling*

Concordance is measured with the Antolini concordance[33] implementation from PyCox with score contributions weighted by Kaplan-Meier estimates.

For Classifier-Cox models, the classifier's output is a binary label predicting mortality by a time horizon. We treat this output as a generic biomarker and fit a Cox regression on validation data with the scikit-survival package[34-36]. This fitting step includes the full[time-to-event, event] information, so these regressions still model mortality over the full dataset's timespan.

The LH, MTLR, and DeepHit models fit survival functions at discrete time points. We used 100 uniformly distributed time points between '0' and the maximum time-to-event in each validation dataset. DeepSurv models and Cox regressions can be queried at any time of interest.

Before measuring model performance, to maintain consistency, we sample the Cox and DeepSurv survival functions at the same 100 time points used to fit the LH, MTLR, and DeepHit models. This sampling greatly simplifies concordance calculations, which would otherwise need survival estimates for every ECG at every unique time-to-event.

When measuring model performance at the 1,2,5,10-year marks, we use the temporally closest sampled survival function value (e.g. MIMIC-IV data has a maximum follow-up time of almost 13 years; one-year AUROC is measured from the survival function at the 1.008-year mark).

See[37] for a thorough guide to survival modeling.

### *Data labeling*

When training classifiers, we label events as a '1' when patients experience an event by horizon *h*, else 0, even if censored. This same labeling applied to AUROC/PRC measures. While there is no standard approach for integrating censored patients with time-dependent measures[38], past work tends to exclude patients that have not reached the horizon-of-interest[4,5], which inflates the event rate (consider a 1000-person study with an annual 90 percent drop-out rate and 1 percent event rate: after two years there are ~5 events and ~10 controls). We chose to mark censored patients as non-events for classifier labels and AUPRC to avoid inflating the event rate. This is reasonable for Code-15, where drop-out rates exceed event rates, and for MIMIC-IV and BCH, where censoring implies lower patient risk. While such choices cannot be avoided in measuring AUPRC/ROC, Deep-Survival methods inherently avoid them.

## SUPPLEMENTAL FIGURES

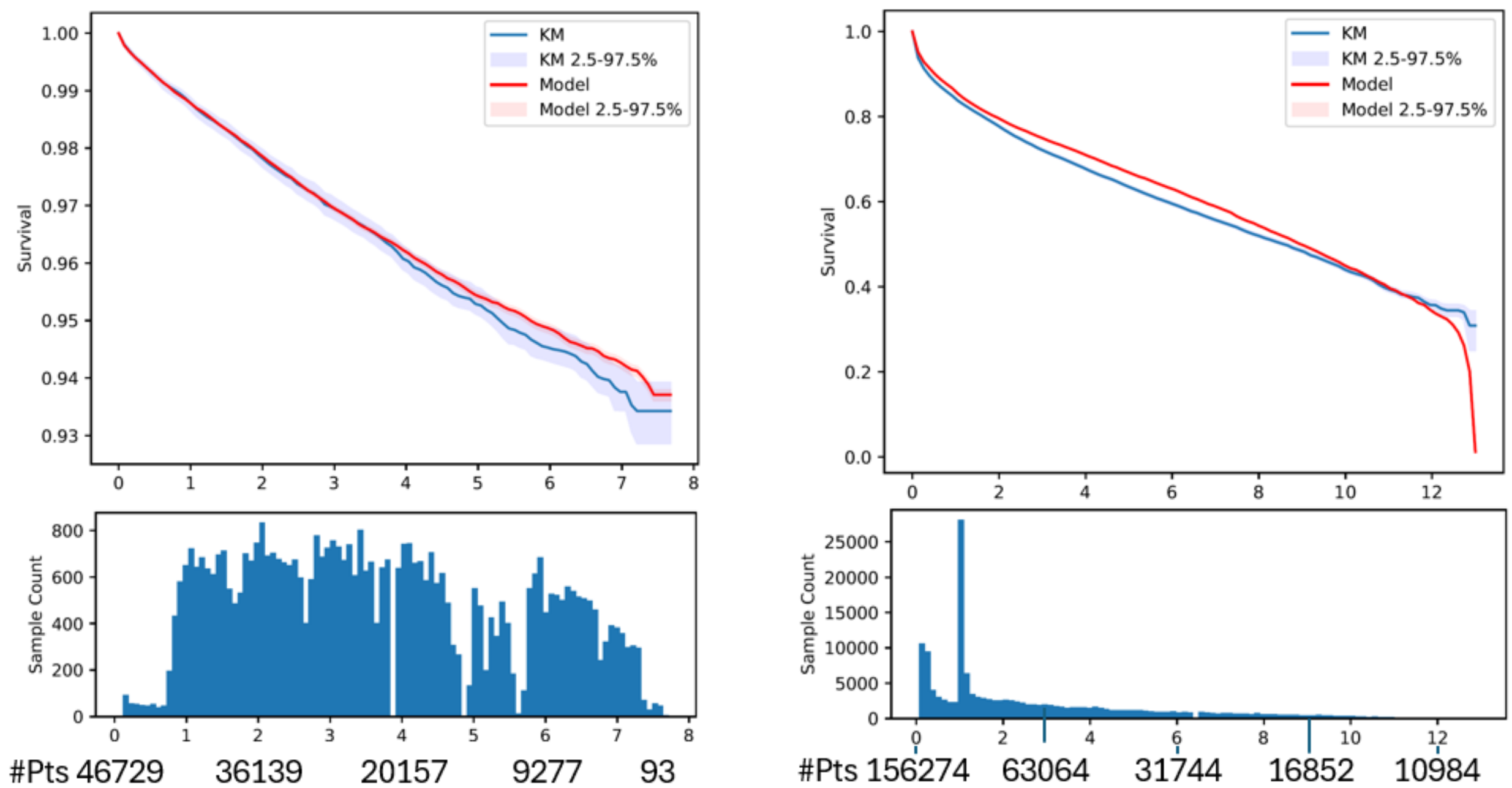

(a) Code-15 (b) MIMIC-IV

**Figure S1.** Kaplan-Meier and population-averaged survival functions for the highest-concordance-index non-demographic Code-15 (left) and MIMIC-IV (right) models. Bottom subplots show sample count at each time point. Plots show median and 95% confidence intervals per time point (100) over 100 bootstraps. Red values indicate the number of patients under observation. Note differences in scale.

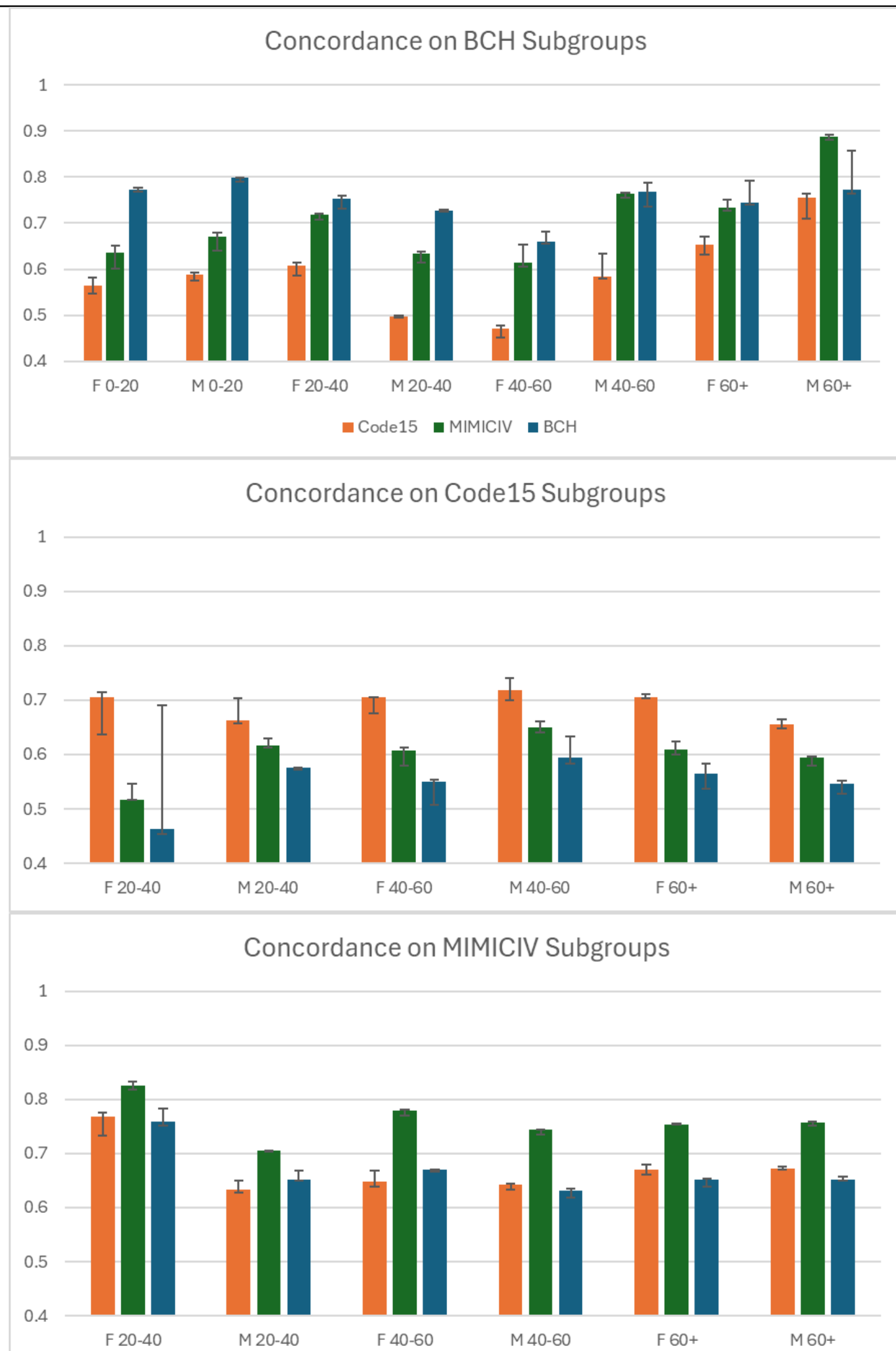

**Figure S2.** Median (IQR) model concordance per patient subgroup, evaluating best model with demographics. Bar color indicates the training site used. Code-15 and MIMIC-IV Age 0-20 are not shown due to low event count; for reference, see Table S2. Models perform best on their own test set, except possibly MIMIC-IV on BCH M60+.

## SUPPLEMENTAL TABLES

**Table S1: Median concordance (Inter-quartile range) for locally trained models, grouped by dataset, survival approach, model inputs, neural network architecture. Values in bold denote the best performing models per dataset.**

| | | | Survival Approach | | | | | | | |
|---|---|---|---|---|---|---|---|---|---|---|
| **Dataset** | **Model Inputs** | **Model Architecture** | **Cla-01** | **Cla-02** | **Cla-05** | **Cla-10** | **DeepHit** | **DeepSurv** | **LH** | **MTLR** |
| **BCH** | **y ~ Age + Sex** | **Feedforward Net** | 0.60 (0.60-0.61) | 0.61 (0.61-0.61) | 0.62 (0.62-0.63) | 0.64 (0.64-0.64) | 0.65 (0.65-0.66) | 0.65 (0.65-0.65) | 0.63 (0.63-0.64) | 0.64 (0.63-0.64) |
| | | **XGB** | 0.56 (0.56-0.57) | 0.59 (0.58-0.59) | 0.62 (0.61-0.62) | 0.63 (0.63-0.63) | - | - | - | - |
| | **y ~ ECG** | **ECGTransForm** | 0.68 (0.67-0.68) | 0.67 (0.67-0.67) | 0.64 (0.59-0.67) | 0.66 (0.60-0.68) | 0.30 (0.11-0.46) | 0.76 (0.67-0.77) | 0.09 (0.01-0.11) | 0.00 (0.00-0.00) |
| | | **InceptionTime** | 0.77 (0.76-0.77) | 0.77 (0.77-0.77) | 0.78 (0.77-0.78) | 0.78 (0.78-0.79) | 0.78 (0.77-0.78) | **0.79 (0.78-0.79)** | 0.78 (0.76-0.78) | 0.78 (0.78-0.79) |
| | | **ResNet** | 0.78 (0.77-0.78) | 0.76 (0.76-0.77) | 0.78 (0.78-0.79) | 0.78 (0.78-0.78) | 0.77 (0.77-0.79) | **0.79 (0.77-0.79)** | 0.78 (0.78-0.79) | **0.79 (0.79-0.80)** |
| | **y ~ ECG + Age + Sex** | **ECGTransForm** | 0.59 (0.59-0.60) | 0.62 (0.61-0.62) | 0.63 (0.63-0.63) | 0.65 (0.64-0.74) | 0.66 (0.64-0.66) | 0.65 (0.65-0.65) | 0.64 (0.64-0.65) | 0.62 (0.61-0.63) |
| | | **InceptionTime** | 0.78 (0.77-0.78) | 0.78 (0.78-0.79) | **0.79 (0.79-0.79)** | **0.79 (0.79-0.79)** | **0.79 (0.79-0.80)** | **0.79 (0.79-0.80)** | **0.79 (0.79-0.79)** | **0.79 (0.79-0.79)** |
| | | **ResNet** | 0.78 (0.77-0.78) | 0.77 (0.76-0.78) | 0.78 (0.76-0.78) | 0.78 (0.77-0.80) | **0.79 (0.79-0.80)** | **0.79 (0.79-0.80)** | 0.78 (0.77-0.78) | **0.79 (0.79-0.79)** |
| **Code-15** | **y ~ Age + Sex** | **Feedforward Net** | 0.79 (0.79-0.79) | 0.79 (0.79-0.79) | 0.79 (0.79-0.79) | 0.79 (0.79-0.79) | 0.79 (0.79-0.79) | 0.79 (0.79-0.79) | 0.79 (0.79-0.79) | 0.79 (0.79-0.79) |
| | | **XGB** | 0.78 (0.78-0.78) | 0.79 (0.79-0.79) | 0.79 (0.79-0.79) | 0.79 (0.79-0.79) | - | - | - | - |
| | **y ~ ECG** | **ECGTransForm** | 0.25 (0.13-0.38) | 0.54 (0.54-0.54) | 0.50 (0.50-0.50) | - | 0.00 (0.00-0.01) | 0.37 (0.24-0.68) | 0.05 (0.01-0.19) | 0.00 (0.00-0.00) |
| | | **InceptionTime** | 0.75 (0.75-0.75) | 0.77 (0.77-0.78) | 0.78 (0.78-0.78) | 0.78 (0.78-0.78) | 0.79 (0.78-0.79) | 0.79 (0.78-0.79) | 0.78 (0.78-0.79) | 0.79 (0.79-0.79) |
| | | **ResNet** | 0.77 (0.76-0.77) | 0.77 (0.77-0.78) | 0.76 (0.75-0.79) | 0.77 (0.77-0.78) | 0.80 (0.80-0.80) | 0.80 (0.80-0.80) | 0.80 (0.79-0.80) | 0.80 (0.79-0.80) |
| | **y ~ ECG + Age + Sex** | **ECGTransForm** | 0.79 (0.79-0.79) | 0.79 (0.79-0.79) | 0.79 (0.79-0.79) | 0.79 (0.79-0.79) | 0.79 (0.79-0.79) | 0.79 (0.79-0.79) | 0.79 (0.79-0.79) | 0.79 (0.79-0.79) |
| | | **InceptionTime** | 0.81 (0.81-0.81) | 0.81 (0.81-0.82) | 0.82 (0.82-0.82) | 0.82 (0.82-0.82) | 0.82 (0.82-0.82) | 0.82 (0.82-0.82) | 0.82 (0.82-0.82) | 0.82 (0.82-0.82) |
| | | **ResNet** | 0.80 (0.80-0.81) | 0.81 (0.81-0.82) | 0.81 (0.81-0.82) | 0.82 (0.81-0.82) | 0.82 (0.81-0.82) | 0.82 (0.81-0.83) | 0.81 (0.81-0.82) | **0.83 (0.82-0.83)** |
| **MIMIC-IV** | **y ~ Age + Sex** | **Feedforward Net** | 0.66 (0.66-0.66) | 0.66 (0.66-0.66) | 0.66 (0.66-0.66) | 0.66 (0.66-0.66) | 0.65 (0.65-0.65) | 0.65 (0.65-0.65) | 0.65 (0.65-0.65) | 0.65 (0.65-0.65) |
| | | **XGB** | 0.65 (0.65-0.66) | 0.66 (0.65-0.66) | 0.66 (0.66-0.66) | 0.66 (0.66-0.66) | - | - | - | - |
| | **y ~ ECG** | **ECGTransForm** | 0.00 (0.00-0.32) | 0.23 (0.01-0.47) | 0.00 (0.00-0.00) | 0.00 (0.00-0.27) | 0.00 (0.00-0.00) | - | 0.73 (0.00-0.73) | 0.00 (0.00-0.00) |
| | | **InceptionTime** | 0.76 (0.76-0.76) | 0.76 (0.76-0.76) | 0.76 (0.76-0.76) | 0.75 (0.75-0.75) | 0.77 (0.76-0.77) | 0.76 (0.76-0.76) | 0.76 (0.76-0.76) | 0.76 (0.76-0.76) |
| | | **ResNet** | 0.77 (0.76-0.77) | 0.76 (0.76-0.76) | 0.76 (0.76-0.76) | 0.75 (0.75-0.76) | 0.77 (0.77-0.77) | 0.75 (0.75-0.76) | 0.77 (0.76-0.77) | 0.77 (0.76-0.77) |
| | **y ~ ECG + Age + Sex** | **ECGTransForm** | 0.66 (0.66-0.66) | 0.66 (0.66-0.67) | 0.66 (0.66-0.66) | 0.66 (0.66-0.66) | 0.65 (0.65-0.65) | - | 0.65 (0.65-0.65) | 0.65 (0.65-0.65) |
| | | **Feedforward Net** | 0.71 (0.68-0.71) | 0.72 (0.69-0.72) | 0.69 (0.69-0.69) | 0.69 (0.68-0.70) | 0.68 (0.65-0.72) | - | 0.71 (0.69-0.72) | 0.64 (0.64-0.64) |
| | | **InceptionTime** | 0.77 (0.77-0.78) | 0.77 (0.77-0.77) | 0.77 (0.77-0.77) | 0.76 (0.76-0.76) | **0.78 (0.78-0.78)** | 0.77 (0.77-0.77) | **0.78 (0.78-0.78)** | 0.77 (0.77-0.78) |
| | | **ResNet** | **0.78 (0.77-0.78)** | 0.77 (0.77-0.77) | 0.77 (0.77-0.77) | 0.76 (0.76-0.76) | **0.78 (0.77-0.78)** | 0.77 (0.76-0.77) | **0.78 (0.77-0.78)** | **0.78 (0.77-0.78)** |
| | | **XGB** | 0.73 (0.73-0.73) | 0.73 (0.73-0.73) | 0.73 (0.73-0.73) | 0.72 (0.72-0.72) | - | - | - | - |
| | **y ~ ECG + Age + Sex + Machine Measures** | **ECGTransForm** | 0.72 (0.72-0.72) | 0.72 (0.69-0.72) | 0.69 (0.69-0.71) | 0.69 (0.69-0.69) | 0.72 (0.72-0.72) | - | 0.72 (0.72-0.72) | 0.68 (0.67-0.69) |
| | | **InceptionTime** | 0.77 (0.77-0.77) | 0.77 (0.77-0.77) | 0.76 (0.76-0.76) | 0.75 (0.75-0.76) | 0.77 (0.77-0.77) | - | 0.77 (0.77-0.77) | 0.77 (0.77-0.77) |
| | | **ResNet** | 0.77 (0.77-0.77) | 0.77 (0.77-0.77) | 0.76 (0.76-0.77) | 0.76 (0.76-0.76) | 0.77 (0.77-0.77) | - | 0.77 (0.77-0.77) | 0.77 (0.76-0.77) |

**Table S2: (Top) Mortality events and rates by age and sex in each test cohort. (Bottom) Mortality events and rates associated with a mortality by year *h*.**

| | | Mortality Events (%) by Age and Sex, Test Set | | |
|---|---|---|---|---|
| **Age** | **Sex** | **BCH** | **MIMIC-IV** | **Code-15** |
| [0,20) | F | 727 (4.2) | 5 (1.4) | 2 (0.2) |
| | M | 994 (4.9) | 2 (0.7) | 3 (0.3) |
| [20,40) | F | 307 (10.2) | 441 (5.3) | 29 (0.3) |
| | M | 463 (16.1) | 542 (8.4) | 37 (0.7) |
| [40,60) | F | 110 (15.5) | 2699 (13.7) | 96 (1.1) |
| | M | 127 (19.2) | 4290 (19.8) | 130 (2.5) |
| <60 | F | 1144 (5.4) | 3145 (11.1) | 127 (0.7) |
| | M | 1584 (6.6) | 4834 (17) | 170 (1.5) |
| 60+ | F | 59 (28.8) | 16773 (34.2) | 651 (6.6) |
| | M | 20 (17.2) | 18705 (37.1) | 740 (9.8) |
| All | F | 1203 (5.6) | 19918 (25.7) | 778 (2.8) |
| | M | 1604 (6.7) | 23539 (29.8) | 910 (4.8) |
| | Both | 2807 (6.2) | 43457 (27.8) | 1688 (3.6) |

| | Mortality Events (%) by year | | |
|---|---|---|---|
| **Year** | **BCH** | **MIMIC-IV** | **Code-15** |
| 1 | 1638 (0.9) | 116185 (14.8) | 2804 (1.2) |
| 2 | 2730 (1.5) | 144446 (18.4) | 4907 (2.1) |
| 5 | 5641 (3.1) | 192334 (24.5) | 7944 (3.4) |
| 10 | 8735 (4.8) | 295173 (37.6) | 8411 (3.6) |
| Max | 12556 (6.9) | 218240 (27.8) | 8411 (3.6) |

**Table S3. Deep-Survival outperform Classifier-Cox Models.** Unpaired Mann-Whitney U Test; per row: 4 survival models, 2 CNNS, 5 random seeds. Grey rows indicate matching Test, Train datasets.

| Train | Test | Type | ECG Only Concordance Median (IQR) | P | ECG + Age/Sex Concordance Median (IQR) | P |
|---|---|---|---|---|---|---|
| Code-15 | Code15 | CC | 0.774 (0.756, 0.780) | **4.65E-10** | 0.815 (0.808, 0.818) | **2.33E-04** |
| | | DS | 0.791 (0.785, 0.799) | | 0.819 (0.817, 0.821) | |
| | BCH | CC | 0.710 (0.702, 0.723) | 4.68E-01 | 0.734 (0.725, 0.742) | **3.81E-02** |
| | | DS | 0.709 (0.654, 0.731) | | 0.724 (0.638, 0.743) | |
| | MIMICIV | CC | 0.662 (0.650, 0.678) | 4.09E-02 | 0.689 (0.682, 0.701) | 6.13E-01 |
| | | DS | 0.673 (0.659, 0.689) | | 0.691 (0.684, 0.700) | |
| MIMIC-IV | Code15 | CC | 0.642 (0.632, 0.654) | 4.33E-01 | 0.684 (0.676, 0.697) | **1.25E-02** |
| | | DS | 0.645 (0.632, 0.660) | | 0.698 (0.683, 0.723) | |
| | BCH | CC | 0.699 (0.674, 0.713) | **6.13E-05** | 0.738 (0.711, 0.757) | **1.95E-06** |
| | | DS | 0.666 (0.656, 0.684) | | 0.690 (0.653, 0.718) | |
| | MIMICIV | CC | 0.761 (0.756, 0.763) | **6.56E-03** | 0.771 (0.763, 0.775) | **1.98E-03** |
| | | DS | 0.764 (0.761, 0.766) | | 0.775 (0.770, 0.777) | |
| BCH | Code15 | CC | 0.576 (0.556, 0.586) | 1.42E-01 | 0.616 (0.603, 0.635) | **3.31E-02** |
| | | DS | 0.564 (0.548, 0.581) | | 0.653 (0.599, 0.678) | |
| | BCH | CC | 0.775 (0.769, 0.782) | **2.07E-02** | 0.782 (0.774, 0.789) | **2.56E-05** |
| | | DS | 0.783 (0.773, 0.789) | | 0.792 (0.788, 0.795) | |
| | MIMICIV | CC | 0.602 (0.568, 0.627) | 7.84E-01 | 0.661 (0.634, 0.673) | **2.48E-03** |
| | | DS | 0.595 (0.568, 0.620) | | 0.677 (0.663, 0.683) | |

**Table S4: Statistical Comparison of ResNet vs InceptionTime models across datasets and choice of covariates.** ResNet models generally outperform InceptionTime models. IT – InceptionTime. RN – ResNet. Wilcoxon Paired Test; per row: 8 survival models, 2 CNNs, 5 random seeds. Grey rows indicate matching Test, Train datasets.

| | | | **Covariates** | | | |
|---|---|---|---|---|---|---|
| | | | **y ~ ECG** | | **y ~ ECG + Age + Sex** | |
| **Train** | **Test** | **Architecture** | **Median Concordance (IQR)** | **P** | **Median Concordance (IQR)** | **P** |
| Code15 | Code15 | IT | 0.782 (0.777 - 0.786) | **1.25E-02** | 0.817 (0.815 - 0.820) | 4.52E-01 |
| | | RN | 0.787 (0.773 - 0.800) | | 0.817 (0.810 - 0.821) | |
| | BCH | IT | 0.702 (0.692 - 0.709) | **2.29E-09** | 0.732 (0.720 - 0.736) | 5.10E-01 |
| | | RN | 0.725 (0.711 - 0.736) | | 0.732 (0.694 - 0.747) | |
| | MIMIC-IV | IT | 0.656 (0.646 - 0.662) | **1.81E-08** | 0.686 (0.682 - 0.690) | **9.60E-06** |
| | | RN | 0.687 (0.676 - 0.693) | | 0.701 (0.690 - 0.711) | |
| MIMIC-IV | Code15 | IT | 0.641 (0.632 - 0.651) | 1.32E-01 | 0.695 (0.680 - 0.710) | 2.32E-01 |
| | | RN | 0.646 (0.632 - 0.661) | | 0.686 (0.673 - 0.701) | |
| | BCH | IT | 0.667 (0.658 - 0.682) | **1.20E-08** | 0.702 (0.672 - 0.727) | **2.51E-03** |
| | | RN | 0.700 (0.678 - 0.714) | | 0.732 (0.696 - 0.755) | |
| | MIMIC-IV | IT | 0.761 (0.758 - 0.764) | **2.00E-02** | 0.773 (0.768 - 0.777) | 9.00E-01 |
| | | RN | 0.763 (0.760 - 0.766) | | 0.772 (0.764 - 0.777) | |
| BCH | Code15 | IT | 0.564 (0.551 - 0.578) | 3.07E-01 | 0.635 (0.613 - 0.662) | **4.87E-03** |
| | | RN | 0.577 (0.551 - 0.589) | | 0.614 (0.588 - 0.635) | |
| | BCH | IT | 0.778 (0.770 - 0.784) | 8.17E-02 | 0.790 (0.784 - 0.794) | **1.17E-03** |
| | | RN | 0.781 (0.773 - 0.790) | | 0.783 (0.773 - 0.792) | |
| | MIMIC-IV | IT | 0.571 (0.559 - 0.588) | **1.58E-08** | 0.660 (0.635 - 0.668) | **5.85E-05** |
| | | RN | 0.624 (0.612 - 0.637) | | 0.679 (0.664 - 0.688) | |

**Table S5: Including demographics improves concordance.** Wilcoxon Paired Test; per row: 8 survival models, 2 CNNs, 5 Random seeds. Grey rows indicate matching Test, Train datasets.

| Train | Test | ECG Only Concordance Median (IQR) | ECG + Age + Sex Concordance Median (IQR) | P |
|---|---|---|---|---|
| Code-15 | Code15 | 0.783 (0.773 - 0.793) | 0.817 (0.813 - 0.820) | **7.85E-15** |
| | BCH | 0.709 (0.697 - 0.726) | 0.732 (0.712 - 0.743) | **4.75E-07** |
| | MIMICIV | 0.669 (0.654 - 0.687) | 0.690 (0.683 - 0.700) | **3.96E-13** |
| MIMIC-IV | Code15 | 0.644 (0.632 - 0.656) | 0.690 (0.677 - 0.704) | **1.17E-13** |
| | BCH | 0.680 (0.664 - 0.705) | 0.715 (0.685 - 0.744) | **2.00E-08** |
| | MIMICIV | 0.762 (0.758 - 0.765) | 0.772 (0.767 - 0.777) | **1.43E-14** |
| BCH | Code15 | 0.569 (0.551 - 0.585) | 0.625 (0.602 - 0.657) | **7.58E-14** |
| | BCH | 0.779 (0.771 - 0.787) | 0.789 (0.780 - 0.794) | **2.05E-04** |
| | MIMICIV | 0.598 (0.568 - 0.626) | 0.668 (0.656 - 0.683) | **9.13E-15** |

**Table S1: Median concordance (Inter-quartile range) for local versus globally trained models, grouped by test dataset and survival approach. Results are shown for ResNet architecture, trained with demographic inputs.**

| | | Test Data | | |
|---|---|---|---|---|
| **Survival Approach** | **Train Data** | **Code-15** | **MIMIC-IV** | **BCH** |
| **Cla-01** | **Global** | 0.81 (0.81-0.82) | 0.78 (0.78-0.79) | 0.80 (0.78-0.80) |
| | **Local** | 0.68 (0.63-0.80) | 0.70 (0.68-0.77) | 0.75 (0.74-0.76) |
| **Cla-02** | **Global** | 0.81 (0.81-0.81) | 0.78 (0.78-0.79) | 0.80 (0.79-0.80) |
| | **Local** | 0.68 (0.64-0.81) | 0.70 (0.68-0.77) | 0.76 (0.74-0.76) |
| **Cla-05** | **Global** | 0.82 (0.82-0.82) | 0.78 (0.78-0.78) | 0.80 (0.80-0.80) |
| | **Local** | 0.68 (0.62-0.80) | 0.70 (0.68-0.77) | 0.76 (0.72-0.77) |
| **Cla-10** | **Global** | 0.82 (0.81-0.82) | 0.77 (0.77-0.77) | 0.81 (0.81-0.81) |
| | **Local** | 0.67 (0.60-0.80) | 0.70 (0.68-0.76) | 0.75 (0.72-0.78) |
| **DeepHit** | **Global** | 0.83 (0.83-0.83) | 0.78 (0.78-0.78) | 0.80 (0.80-0.81) |
| | **Local** | 0.73 (0.67-0.81) | 0.70 (0.69-0.77) | 0.69 (0.59-0.79) |
| **DeepSurv** | **Global** | 0.81 (0.81-0.81) | 0.77 (0.77-0.78) | 0.80 (0.80-0.80) |
| | **Local** | 0.70 (0.68-0.81) | 0.70 (0.69-0.76) | 0.76 (0.73-0.79) |
| **LH** | **Global** | 0.83 (0.82-0.83) | 0.78 (0.78-0.78) | 0.79 (0.78-0.80) |
| | **Local** | 0.70 (0.58-0.81) | 0.70 (0.68-0.77) | 0.74 (0.73-0.77) |
| **MTLR** | **Global** | 0.82 (0.82-0.82) | 0.78 (0.78-0.78) | 0.80 (0.80-0.80) |
| | **Local** | 0.69 (0.63-0.82) | 0.71 (0.67-0.77) | 0.70 (0.67-0.78) |